\documentclass[published]{JHEP3} % 10pt is ignored!

\JHEP{00(2011)000}

\JHEPspecialurl{http://jhep.sissa.it/JOURNAL/JHEP3.tar.gz}

\usepackage{epsfig,multicol}

\newcommand\fverb{\setbox\pippobox=\hbox\bgroup\verb}
\newcommand\fverbdo{\egroup\medskip\noindent%
            \fbox{\unhbox\pippobox}\ }
\newcommand\fverbit{\egroup\item[\fbox{\unhbox\pippobox}]}
\newbox\pippobox
\title{Schwinger Pair Production in Solitonic Gauge Fields}

\author{Sang Pyo Kim\\
 Department of Physics, Kunsan National University, Kunsan 573-701,
 Korea\\Instituto de F\'{i}sica y Matem\'{a}ticas, Universidad Michoacana de San Nicol\'{a}s de Hidalgo, Apdo. Postal 2-82, C.P. 58040, Morelia, Michoac\'{a}n, Mexico\\ E-mail: \email{sangkim@kunsan.ac.kr}}

\received{}        %%
\revised{}
\accepted{}        %% These are for published papers.

\preprint{}  % OR: \preprint{Aaaa/Mm/Yy\\Aaa-aa/Nnnnnn}
\abstract{We study the time-dependent solitonic gauge fields in scalar QED,
in which a charged particle has the energy of reflectionless P\"{o}sch-Teller potential with natural quantum numbers.
Solving the quantum master equation for quadratic correlation functions,
we find the exact pair-production rates as polynomials of inverse square of hyperbolic cosine,
which exhibit solitonic characteristics of a finite total pair production per unit volume and a non-oscillatory behavior
for the entire period, and an exponentially decaying factor in asymptotic regions. It is shown that
the solitonic gauge fields are the simplest solutions of the quantum master equation and
that the back-reaction of the produced pairs does not destabilize the solitonic gauge fields.}


\begin{document}

\section{Introduction}\label{introduction}

Schwinger showed that a constant electric field could produce pairs of charged fermions
in quantum electrodynamics (QED) and of charged bosons in scalar QED \cite{Schwinger}.
Parker showed that an expanding spacetime could also produce particle pairs with opposite quantum numbers \cite{Parker}.
Under background gauge fields or curved spacetimes the out-vacuum can differ from
the in-vacuum and the nontrivial Bogoliugov transformation between two vacua
leads to pairs of the out-vacuum particles in the in-vacuum or vice versa
\cite{DeWitt,DeWitt-book}.
The recent proposals to directly test Schwinger pair production and vacuum polarization effects in ultra-strong laser sources
\cite{Ringwald,MarklundShukla,Dunne09,BPRRSSS,Tajima} and possible astrophysical applications \cite{RVX}
have boosted studies on vacuum instability and pair production \cite{SGD08,HAG08,HADG09,PLMK09,DGS09,BMNNP,DumluDunne10,Dumlu10,HAG10,OHA11,DumluDunne11,HAG11}.
Several field theoretical methods have been elaborated to calculate the pair-production rate in various electric field backgrounds
\cite{GavrilovGitman96,KimPage02,DunneSchubert05,KimPage06,DWGS,KimPage07,GGT,DumluDunne11-2} (for review
and references, see refs. \cite{Dunne-rev,RVX}).
The author and Schubert have recently introduced a time-dependent solitonic gauge field \cite{KimSchubert},
which relates the quantum master equation for quadratic correlation functions to
the KdV equation, whose soliton solution determines the pair-production rate and the other correlation functions.
One interesting feature of the solitonic gauge field is that the pair-production rate is determined by a single hyperbolic function
for the entire period, which exponentially increases from and decreases back to zero in a symmetric way
without any oscillatory behavior throughout the evolution.

In this paper we find a more general class of the so-called ``solitonic gauge fields'' that lead to the
pair-production rate and other correlation functions having the ``solitonic characteristics'' not in the
sense of traveling solitons for the KdV equation
but in sense of finiteness of total number of pairs
produced for the entire period per unit volume, an exponentially decaying behavior in asymptotic regions
and a non-oscillatory tailing behavior. Further the back-reaction of produced pairs should not destabilize
the solitonic gauge fields. These solitonic gauge fields have the form of P\"{o}sch-Teller potential in quantum mechanics, which
also provides the reflectionless potentials whose bound solutions
lead to solitons for the KdV equation \cite{DrazinJohnson}. A massive scalar field
in the global geometry of de Sitter spaces has the P\"{o}sch-Teller potential
for each spherical harmonics, in which particles are produced only in even dimensions
but not in odd dimensions \cite{Mottola,BMS,Polyakov08,Kim10-dS}.
There the pair-production rate, which is obtained by comparing the out-vacuum in the future
infinity with the in-vacuum in the past infinity, is an asymptotic result and
does not give any kinematic information on particle production during the evolution. It is
argued that pair production in de Sitter spaces may cause the decay of the vacuum energy
\cite{Polyakov08,Polyakov10,KrotovPolyakov}.

We show that a charged boson with the P\"{o}sch-Teller potential
for the squared energy,
\begin{eqnarray}
\omega^2_{\bf k} (t) = \omega^2_0 \Bigl( l^2 + \frac{p(p+1)}{\cosh^2 (\omega_0 t)} \Bigr) \label{posch-teller}
\end{eqnarray}
with the free energy $\omega_0$, has the solitonic pair-production rate
for a natural number $p$, which is a consequence of the reflectionless scattering.
The gauge field strength is determined by the parameter $p$. The massive scalar field
in the global covering of de Sitter spaces has the correspondence of the parameters
\begin{eqnarray}
\omega_0 = H, \quad l^2 = \Bigl(\frac{m}{H} \Bigr)^2 - \Bigl(\frac{d}{2} \Bigr)^2, \quad p(p+1) = \lambda +
\frac{d}{2}\Bigl(\frac{d}{2}-1 \Bigr)
\end{eqnarray}
where $H$ is the Hubble constant, $d$ is the spatial dimensions, and $- \lambda$ is the eigenvalue the Laplace operator
on $S^{d}$. The main difference of de Sitter space from QED is that the mass of
the particle is the free parameter while all the other parameters, such as the Hubble constant,
the spatial dimensions and the eigenvalues of
the Laplace operator, are predetermined. As the parameter $p$ can be controlled as the interaction
strength in QED, it is legitimate to call the corresponding fields as the ``solitonic gauge fields,''
though all the results of this paper hold directly for de Sitter spaces.

The method to be employed for finding the pair-production rate is quantum kinematics for
the correlation functions of the spectrum generating algebra of the annihilation and
creation operators for the charged bosons in a gauge field background. The Klein-Gordon
equation in the gauge field has the $SU(1,1)$ algebra \cite{KimSchubert}. This means that
the correlation functions carry the same information as the mode solutions or the propagator
of the Klein-Gordon equation. Though the pair-production rate may be found
from the mode solutions, which is the case of solitonic gauge fields,
the advantage of quantum kinematic approach is that the production-rate
and the correlation functions for quadratic annihilation and creation operators
obey a system of linear differential equations with time-dependent coefficients, which can be integrated
from the initial moment in which no particle is present
and the gauge field vanishes asymptotically.

Solving the quantum master equation, we find the exact pair-production rate in the background (\ref{posch-teller}),
which exhibits the solitonic characteristics. The solitonic solutions may shed light on
two important questions in particle production: the local nature of and the oscillatory behavior of pair production.
The effective action carries both the global and the local information of the background fields
either through the eigenvalues of the field equation or through the real-time propagator \cite{DunneHall}
or through the Bogoliubov coefficients
from the asymptotic mode solution \cite{KLY08,KLY10,Kim11}.
In the in-out formalism the vacuum persistence (twice of the imaginary part of the effective action) relates to the pair-production rate
at one-loop \cite{DeWitt,DeWitt-book}
\begin{eqnarray}
2 {\rm Im} ({\cal L}^{(1)}_{\rm eff} (t)) = \sum_{\bf K} \ln (1+ {\cal N}_{\bf K} (t)),
\end{eqnarray}
where ${\bf K}$ denotes all quantum numbers.
Therefore a physically interesting question may be raised where and when to observe produced pairs. Another surprising aspect
is that though the Schwinger formula in a constant electric field
is constant \cite{Schwinger}, the pair-production rate in general time-dependent electric fields shows
the oscillatory tailing behavior as explained in detailed below.

We also advance an algebraic method to directly solve the quantum master equation. The dependent variables
for the quantum master equation are the quadratic correlation functions of annihilation and creation
operators, which are nonlinear functionals of the complex mode solution, the first derivative,
and their complex conjugates. Therefore the quantum master linear equation, equivalent to
the first order master equation for three independent correlation functions, is a highly
nonlinear functional of the mode solutions \cite{KimSchubert}. However, in terms of the new variable
the master linear equation takes a relatively simple form with the coefficients of
the gauge field and its derivative, which allows an algebraic approach to
the quantum kinematics.  Under the condition that
the square of the first derivative of the profile function for the gauge field is a polynomial of
the profile function itself, we are able to find the recurrence relation for the power series solution for the
quantum master equation. This algebraic method does not assume the profile of the gauge field {\it a priori} but determines
the profile function from the requirement that the recurrence relation should terminate for soliton solutions.
Remarkably we find that the P\"{o}sch-Teller potential (\ref{posch-teller}) with natural numbers $p$
are the simplest solitonic gauge fields from the algebraic method. However, the algebraic method leaves open for other
solitonic gauge fields.

The solitonic gauge fields (natural number $p = n$) strongly contrast with other non-solitonic
gauge fields (non-natural number $p$) of the same potential shape which predict
the oscillatory tailing behavior. The oscillatory behavior may be understood from the quantum master equation
in the non-adiabatic approach \cite{KimSchubert}
\begin{eqnarray}
\frac{d}{dt} (1+ 2 {\cal N}_{\bf k} (t)) &=& \frac{\omega^2_{\bf k} (t)-\omega^2_{\bf k} (t_0)}{\omega^2_{\bf k} (t_0)}  \int_{t_0}^{t} d \xi \Bigl[ (\omega^2_{\bf k} (\xi) -\omega^2_{\bf k} (t_0))
(1+ 2 {\cal N}_{\bf k} (\xi)) \nonumber\\ &&\times \cos \Bigl( \frac{1}{\omega_{\bf k} (t_0)}\int_{\xi}^{t} d \xi' (\omega^2_{\bf k} (\xi') + \omega^2_{\bf k} (t_0))\Bigr) \Bigr], \label{mas-eq}
\end{eqnarray}
or the quantum Vlasov equation from the adiabatic approach \cite{KESCM91,KESCM92,SBRSPT98,KME98,SBRPST99,HAG08}
\begin{eqnarray}
\frac{d}{dt} (1+ 2 {\cal N}_{\bf k} (t)) = \frac{d \omega^2_{\bf k} (t)/dt}{4 \omega^2_{\bf k} (t)}  \int_{t_0}^{t} d \xi \Bigl[
\frac{d \omega^2_{\bf k} (\xi)/d\xi}{\omega^2_{\bf k} (\xi)}
(1+ 2 {\cal N}_{\bf k} (\xi)) \cos \Bigl( 2 \int_{t_0}^{\xi} d \xi' \omega^2_{\bf k} (\xi') \Bigr) \Bigr], \label{ad-mas-eq}
\end{eqnarray}
where $\omega_{\bf k} (t)$ is the energy in the time-dependent electric field and $\omega_{\bf k} (t_0)$ denotes the initial value.
In the asymptotic regions $(t = \pm \infty)$, eq. (\ref{mas-eq}) or (\ref{ad-mas-eq}) weights the envelop of exponentially decreasing function
with the sinusoidal function dominated either by $\cos (2 \omega_{\bf k} (t_0) (t - \xi))$ in the
non-adiabatic approach or by $\cos (2 \omega_{\bf k} (t_0) t)$ in the adiabatic approach.
Therefore the non-oscillatory tailing behavior of solitonic gauge fields is an astonishing result since
the oscillations should exactly cancel.

We investigate the back-reaction of the produced pairs by computing the induced current and
the induced electric field in a one-soliton gauge field. The renormalized
current and thereby the induced electric field has
the leading term proportional to the solitonic gauge field background by a factor of the fine structure constant.
This means that the back-reaction through the induced current and the electric field may be neglected for
the solitonic gauge field. The full scenario for the back-reaction problem is the quantum master
equation in the gauge field background plus the induced gauge field and the equation for the current, which makes
the system coupled and highly nonlinear. We leave it as an open question whether the solitonic gauge fields could keep the
same shape or whether any solitonic gauge field could exist or not.

The organization of this paper is as follows. In sec. \ref{sol-non-sol}
we exploit the properties of solitonic and non-solitonic gauge fields.
In sec. \ref{pair-mod} we find the pair-production rate from the mode solutions.
In sec.  \ref{ser-sol} we find the power series solutions of the quantum master equation
for the solitonic and non-solitonic gauge fields.
In sec. \ref{gen-sol} we solve the quantum master equation by an algebraic method and show
that the solitonic gauge fields are
the simplest one allowing polynomial solutions.
In sec. \ref{pair-sol} we investigate the pair-production rate and correlation functions in solitonic gauge fields.
In sec. \ref{non-sol-pair} we compare pair production in non-solitonic gauge fields
with that in solitonic gauge fields.
In sec. \ref{back-reaction} we study the back-reaction of produced pairs by computing
the induced current and electric field and show that the back-reaction is
always order of the fine structure constant and thus negligible.

\section{Solitonic and Non-Solitonic Gauge Fields}\label{sol-non-sol}

The recently introduced time-dependent solitonic gauge field \cite{KimSchubert} relates the quantum master equation to the KdV equation, whose solution determines the pair-production rate and the correlation functions of pair annihilation and creation. We first generalize it to (i) the solitonic gauge fields
\begin{eqnarray}
q A_{(n)\parallel} (t) = k_{\parallel} - \Bigl( k_{\parallel}^2 + \frac{n(n+1) \omega_0^2}{\cosh^2 (\omega_0 t)} \Bigr)^{1/2},
\quad (n = 1, 2, 3, \cdots), \label{sol-gauge}
\end{eqnarray}
where $q$ and $m$ are the charge and mass and $\omega_0 = \sqrt{{\bf k}^2 + m^2}$ is the free particle energy
in the asymptotic limits $t = \pm \infty$. The case of $n=1$ is related with the KdV equation \cite{KimSchubert}.
For the zero-longitudinal momentum the gauge fields simply become $A_{(n)\parallel} (t) = - \sqrt{n(n+1)}\omega_0/(q\cosh (\omega_0 t))$
and the potential difference is zero between $t = - \infty$ and $t = \infty$,
and the electric field
\begin{eqnarray}
E_{(n)\parallel} (t) = - \frac{\sqrt{n(n+1)}\omega_0^2}{q} \frac{\sinh (\omega_0 t)}{\cosh^2(\omega_0 t)}
\end{eqnarray}
changes the polarity and exponentially decreases in the asymptotic limits $t = \pm \infty$.
The Fourier-component of a charged scalar in the solitonic gauge field obeys the mode equation [in units of $\hbar = c =1$]
\begin{eqnarray}
\ddot{\varphi}_{(n){\bf k}} (t) + \omega_0^2 \Bigl(1 + \frac{n(n+1)}{\cosh^2 (\omega_0 t)} \Bigr)
\varphi_{(n){\bf k}} (t) = 0. \label{mod-eq}
\end{eqnarray}
As the solution depends on the free particle energy,
we rescale the time as $t:= \omega_0 t$ and measure the time in the Compton time with the equivalent mass of relativistic energy
or equivalently set $\omega_0 = 1$ for simplicity, delete the momentum ${\bf k}$
unless necessary, and denote
\begin{eqnarray}
\omega^2_{(n)} (t) := 1 + \frac{n(n+1)}{\cosh^2 t}, \quad (n = 1, 2, 3, \cdots). \label{sol-omega}
\end{eqnarray}
The next generalization is (ii) the two parameter-dependent gauge fields
\begin{eqnarray}
\omega^2_{(pl)} (t) := l^2 + \frac{p(p+1)}{\cosh^2 t}, \quad \quad (p \neq 1, 2, 3, \cdots) \label{non-sol-omega}
\end{eqnarray}
where $l$ is positive but not necessarily natural numbers. In quantum mechanics the potentials of the form
(\ref{sol-omega}) and (\ref{non-sol-omega}) are known as P\"{o}sch-Teller potential. These potentials have a rich structure
and a variety of interpretations for the Klein-Gordon equation, as will be shown below.

The energy (\ref{sol-omega}) or (\ref{non-sol-omega}) with a natural number $p = n$ is known as the reflectionless potential
for the mode equation (\ref{mod-eq}), whose bound solutions for another natural number $l$ generate
the solitons for the KdV equation \cite{DrazinJohnson}. The reflectionless scattering may be understood semiclassically
by computing the complex Hamilton-Jacobi action in the phase-integral method
\cite{FromanFroman,Kim10,DumluDunne10,DumluDunne11}.
Taking into account the four complex turning points, the leading term of the pair-production rate is \cite{Kim10}
\begin{eqnarray}
{\cal N} \approx 4 \cos^2 (\pi \sqrt{n(n+1)}) e^{- 2 l \pi},
\end{eqnarray}
which destructively interferes approximately for large $n$ such that $\sqrt{n(n+1)} \approx (2n+1)/2$ or completely interferes away
by including the Maslov index contribution $\pi/4$ to the action.
The exact solution for the gauge fields (\ref{non-sol-omega}) is given by the hypergeometric function \cite{gr-table}
\begin{eqnarray}
\varphi_{(pl)} (t) = \frac{e^{- il t}}{\sqrt{2l}} (1+ e^{2t})^{p+1} F(p+1-il, p+1; 1-il; -e^{2t}), \label{mod-sol}
\end{eqnarray}
which has the proper asymptotic behavior at $t = - \infty$ for the Gaussian vacuum with the energy $l (\omega_0)$:
\begin{eqnarray}
\varphi_{(pl)} (t) = \frac{e^{- i l t}}{\sqrt{2l}}. \label{mod-asym}
\end{eqnarray}
In the other asymptotic limit $t = \infty$, the mode solution takes another form
\begin{eqnarray}
\varphi_{(pl)} (t) &=& \frac{e^{-i l t}}{\sqrt{2l}} \frac{\Gamma (1-il) \Gamma (-il)}{\Gamma (p+1-il) \Gamma (-p-il)}
(1+ e^{-2t})^{p+1} F(p+1, p+1+il; 1+il; -e^{-2t}) \nonumber\\
&& + \frac{e^{i lt}}{\sqrt{2l}} \frac{\Gamma (1-il) \Gamma (il)}{\Gamma (p+1) \Gamma (-p)}
(1+ e^{-2t})^{p+1} F(p+1-il, p+1; 1-il; -e^{-2t}). \label{asym-sol2}
\end{eqnarray}
The first case (i) is when $p$ is a natural number $n$, for which the negative frequency
vanishes due to $\Gamma (-n) = (-1)^n \times \infty$.  The solution describes a reflectionless scattering over the
potential, so pairs are not produced in the asymptotic region $t = \infty$.
In the second case (ii) of a non-natural number $p$,
the mode solution (\ref{mod-sol}) has both the positive and the negative frequencies in eq. (\ref{asym-sol2}),
implying pair production in the limit $t = \infty$.
The mode solution may be further written in terms of the associated Legendre function \cite{gr-table,as-table}
\begin{eqnarray}
\varphi_{(pl)} (t) = \frac{\Gamma (1-il)}{\sqrt{2}} P^{il}_{p} (- \tanh t). \label{mod-sol-L}
\end{eqnarray}

The energy (\ref{sol-omega}) or (\ref{non-sol-omega})
has a finite number of bound solutions for positive imaginary $l = i \kappa$. The mode solution (\ref{mod-sol}) exponentially
decreases at $t = - \infty$ while its analytical continuation (\ref{asym-sol2}) to $t = \infty$
has the exponentially increasing branch, the first term, and the exponentially decreasing branch,
the second term. The exponentially increasing branch vanishes when $p - \kappa$ is a non-negative integer:
the number of bound solutions is $[p]$, the largest integer no greater than $p$. For the natural number
$p= n$, the bound solutions are the associated Legendre polynomials $P^{l}_{n} (- \tanh t)$ for $l = 1, 2, \cdots, n$, which generate
the n-solitons for the KdV equation \cite{DrazinJohnson}.

\section{Pair Production in Terms of Mode Solution}\label{pair-mod}

In this section we use the mode solution (\ref{mod-sol}) and the other asymptotic form (\ref{asym-sol2}) to calculate
the pair-production rate in the limit $t = \infty$. However, one may still ask
whether the pair production is absolutely zero or decreases to zero in that asymptotic limit
and ask the kinematical information of pair production during interaction, which will be pursued in this paper.

To find the exact quantum states, we may employ the Lewis-Riesenfeld invariant operators \cite{LewisRiesenfeld}
in the Schr\"{o}dinger picture, which
provide the time-dependent creation and annihilation operators for particles and antiparticles
\cite{MMT70,Kim96,FVV98,KimLee00}
\begin{eqnarray}
\hat{a}^{\dagger}_{(pl) {\bf k}} (t) &=& - i [\varphi_{(pl) {\bf k}} (t) \hat{\pi}_{(pl) {\bf k}} - \dot{\varphi}_{(pl) {\bf k}} (t) \hat{\phi}_{(pl) {\bf k}}^{\dagger}], \nonumber\\
\hat{a}_{(pl) {\bf k}} (t) &=& i [\varphi_{(pl) {\bf k}}^* (t) \hat{\pi}_{(pl) {\bf k}}^{\dagger} - \dot{\varphi}_{(pl) {\bf k}}^* (t) \hat{\phi}_{(pl) {\bf k}}],
\label{fock-pa}
\end{eqnarray}
and for antiparticles
\begin{eqnarray}
\hat{b}^{\dagger}_{(pl) - {\bf k}} (t) &=& -i [\varphi_{(pl) {\bf k}} (t) \hat{\pi}_{(pl) {\bf k}}^{\dagger}  - \dot{\varphi}_{(pl) {\bf k}} (t) \hat{\phi}_{(pl) {\bf k}}],\nonumber\\
\hat{b}_{(pl) -{\bf k}} (t) &=& i [\varphi_{(pl) {\bf k}}^* (t) \hat{\pi}_{(pl) {\bf k}} - \dot{\varphi}_{(pl) {\bf k}}^* (t) \hat{\phi}_{(pl) {\bf k}}^{\dagger}]. \label{fock-an}
\end{eqnarray}
Here $\hat{\phi}_{(pl) {\bf k}}$ and $\hat{\pi}_{(pl) {\bf k}}$ are the Fourier component of the field and the momentum,
all being Schr\"{o}dinger operators, and $\varphi_{(pl) {\bf k}}$ is the complex solution (\ref{mod-sol}).
The mode solution (\ref{mod-sol}) satisfies the Wronskian condition from the canonical quantization
\begin{eqnarray}
{\rm Wr} [\varphi_{(pl) {\bf k}}, \varphi^*_{(pl) {\bf k}}] \equiv \varphi_{(pl) {\bf k}} (t) \dot{\varphi}^*_{(pl) {\bf k}} (t)
- \varphi^*_{(pl) {\bf k}} (t) \dot{\varphi}_{(pl) {\bf k}} (t) = i, \label{wr-con}
\end{eqnarray}
which is equivalent to the equal-time commutation relations
\begin{eqnarray}
\lbrack \hat{a}_{(pl) {\bf k}} (t), \hat{a}^{\dagger}_{(pl){\bf k}'} (t) \rbrack &=&
 \lbrack \hat{b}_{(pl) -{\bf k}} (t), \hat{b}^{\dagger}_{(pl) -{\bf k}'} (t) \rbrack = \delta ({\bf k} - {\bf k}').
 \label{comm}
\end{eqnarray}
Then the Bogoliubov transformation between the past infinity $t_0 = - \infty$ and the present time $t$ is given by
\begin{eqnarray}
\hat{a}_{(pl) {\bf k}} (- \infty) &=& \mu_{(pl) {\bf k}} (- \infty, t) \hat{a}_{(pl) {\bf k}} (t) + \nu_{(pl) {\bf k}} (- \infty, t) \hat{b}_{(pl) - {\bf k}}^{\dagger} (t), \nonumber\\
\hat{b}^{\dagger}_{(pl) -{\bf k}} (- \infty) &=& \mu^*_{(pl) {\bf k}} (- \infty, t) \hat{b}^{\dagger}_{(pl) -{\bf k}} (t)
+ \nu^*_{(pl) {\bf k}} (- \infty, t) \hat{a}_{(pl) {\bf k}} (t),\label{bog-tr}
\end{eqnarray}
where
\begin{eqnarray}
\mu_{(pl) {\bf k}} (- \infty, t) &=& i {\rm Wr} [\varphi^*_{(pl) {\bf k}} (- \infty), \varphi_{(pl) {\bf k}} (t)], \nonumber\\
\nu_{(pl) {\bf k}} (- \infty, t) &=& i {\rm Wr} [\varphi^*_{(pl) {\bf k}} (- \infty), \varphi^*_{(pl) {\bf k}} (t)]. \label{bog-co}
\end{eqnarray}
The Wronskian (\ref{wr-con}) makes the Bogoliubov coefficients satisfy the bosonic relation
\begin{eqnarray}
|\mu_{(pl) {\bf k}} (- \infty, t)|^2 - |\nu_{(pl) {\bf k}} (- \infty, t)|^2 = 1. \label{bog-rel}
\end{eqnarray}

The time-dependent ground state $\vert 0_{(pl) {\bf k}}, t \rangle$, a Gaussian state, is nullified by
the annihilation operators for a given ${\bf k}$
\begin{eqnarray}
\hat{a}_{(pl) {\bf k}} (t) \vert 0_{(pl) {\bf k}}, t \rangle =  \hat{b}_{(pl) -{\bf k}} (t) \vert 0_{(pl) {\bf k}}, t \rangle = 0,
\end{eqnarray}
and the time-dependent vacuum is the tensor product
\begin{eqnarray}
\vert 0_{(pl)}, t \rangle = \prod_{\bf k} \vert 0_{(pl) {\bf k}}, t \rangle.
\end{eqnarray}
The mean number of pairs carried by the time-dependent vacuum as measured by the in-vacuum number operator is given by
\begin{eqnarray}
\langle 0_{(pl)}, t \vert \hat{a}^{\dagger}_{(pl) {\bf k}} (- \infty) \hat{a}_{(pl) {\bf k}} (- \infty) \vert 0_{(pl)}, t \rangle
&=& |\nu_{(pl) {\bf k}} (- \infty, t)|^2, \nonumber\\
\langle 0_{(pl)}, t \vert \hat{b}^{\dagger}_{(pl) -{\bf k}} (- \infty) \hat{b}_{(pl) -{\bf k}} (- \infty)
\vert 0_{(pl) }, t \rangle &=& |\nu_{(pl) {\bf k}} (- \infty, t)|^2. \label{pair-ra}
\end{eqnarray}
Similarly, the mean number of pairs contained in the in-vacuum now as measured by the number operator at $t$ is the same as
eq. (\ref{pair-ra}) since $\nu_{(pl) {\bf k}} (t,- \infty)=\nu^*_{(pl) {\bf k}} (- \infty, t)$.
The equal number of particles and antiparticles are produced due to charge neutrality of the in-vacuum and the electric field.
Interpreting as the number of pairs the expectation value of the time-dependent number operator
in the in-vacuum or the number of in-vacuum particles in the time-dependent state
is a subtle issue (see refs. \cite{KME98,HAG08}). In the adiabatic case in which $|\dot{\omega}^2/\omega^4|, |\ddot{\omega}/\omega^3|
\ll 1$, we may use the adiabatic solution to count the adiabatic particle number \cite{BirrelDavies} and the number of pairs
(\ref{pair-ra}) approaches the adiabatic one. The gauge potentials (\ref{sol-omega}) and (\ref{non-sol-omega})
satisfy the adiabaticity, which will be discussed in detail in sec. \ref{dis}.
The pair-production rate (\ref{pair-ra}) from the solution (\ref{mod-sol}) is
\begin{eqnarray}
{\cal N}_{(pl)} (t) = \frac{1}{4l^2} |\dot{A}_{(pl)} (t)|^2, \label{pair-mod-sol}
\end{eqnarray}
where $A_{(pl)}$ is the time-dependent amplitude
\begin{eqnarray}
A_{(pl)} (t) = (1+ e^{2t})^{p+1} F(p+1-il, p+1; 1-il; -e^{2t}). \label{mod-amp}
\end{eqnarray}
In the asymptotic limits $t = \pm \infty$, the pair-production rate for the solitonic gauge field is approximately given by
\begin{eqnarray}
{\cal N}_{(nl)} (t) = \Bigl(\frac{n(n+1)}{2l} \Bigr)^2 \frac{1}{4(1+l^2)} \frac{1}{\cosh^4 t}. \label{leading-pair}
\end{eqnarray}
The pair-production rate in the non-solitonic gauge field still has the same form (\ref{leading-pair}) in the limit $t = - \infty$,
but the other limit $t = \infty$ will be separately treated in sec. \ref{non-sol-pair}.

The physical meaning of $l$ in eq. (\ref{non-sol-omega}) is
the squeezed vacuum of the Minkowski one for $l =1$. The squeezed vacuum is the Gaussian vacuum
with the energy $l (\omega_0)$ while the Minkowski vacuum has the minimum energy $1(\omega_0)$ and the uncertainty
$1/2$ \cite{KimNoz,KimKim99}.
Comparing the asymptotic solution (\ref{mod-asym})
for $l \neq 1$ with that for $l=1$, we find the Bogoliubov coefficients
\begin{eqnarray}
\mu_{(nl)(n)} = i {\rm Wr} [\varphi_{(n)}^* (t), \varphi_{(nl)} (t)] = \frac{l+1}{2 \sqrt{l}} e^{- i (l-1) t}, \nonumber\\
\nu_{(nl)(n)} = i {\rm Wr} [\varphi_{(n)}^* (t), \varphi_{(nl)}^* (t)] = - \frac{l-1}{2 \sqrt{l}} e^{- i (l+1) t}. \label{bog-sq}
\end{eqnarray}
The Bogoliubov transformation (\ref{bog-sq}) implies that the vacuum state for $l$ is the squeezed vacuum of the Minkowski one
in the asymptotic limit $t = - \infty$ \cite{KLY08}
\begin{eqnarray}
\vert 0_{(nl)}, t \rangle &=& \exp \Bigl[r_{l} \Bigl(\hat{a}_{(n)} (t) \hat{b}_{(n)} (t) e^{2it} - \hat{a}^{\dagger}_{(n)} (t)
\hat{b}^{\dagger}_{(n)} (t) e^{-2it} \Bigr) \Bigr] \nonumber\\
&& \times \exp \Bigl[i (l-1) t \Bigl(\hat{a}^{\dagger}_{(n)} (t) \hat{a}_{(n)} (t))+ \hat{b}_{(n)} (t) \hat{b}^{\dagger}_{(n)} (t) \Bigr)
\Bigr] \vert 0_{(n)}, t \rangle,
\end{eqnarray}
where
\begin{eqnarray}
\cosh r_{l} = \frac{l+1}{2 \sqrt{l}}, \quad  \sinh r_{l} = \frac{l-1}{2 \sqrt{l}}.
\end{eqnarray}
The probability for the squeezed vacuum to be in the number state $\vert k_{(n)}, t \rangle$ of the Minkowski vacuum
is
\begin{eqnarray}
P_k = |\langle k_{(n)}, t \vert 0_{(nl)}, t \rangle|^2 = \frac{\tanh^{2k} r_{l}}{\cosh^2 r_{l}} = \frac{4l}{(l+1)^2}
\Bigl(\frac{l-1}{l+1} \Bigr)^{2k}.
\end{eqnarray}
Thus the squeezed vacuum has the mean number of pairs with respect to the Minkowski vacuum and its Fock space
\begin{eqnarray}
{\cal N}_{(nl)(n)} = \sum_{k= 0}^{\infty} k P_k = \frac{(l-1)^2}{4l} = |\nu_{(nl)(n)}|^2.
\end{eqnarray}

\section{Series Solution for Master Equation}\label{ser-sol}

The linear combinations of quadratic annihilation and creation operators for each component of the Klein-Gordon equation, not including interaction terms,
in an electromagnetic field background or in a curved spacetime, constitute the $SU(1,1)$ algebra \cite{KimSchubert}.
The evolution of these generators carry the same information as the mode solutions in sec. \ref{pair-mod}.
The dynamical equation that determines the pair-production rate from  an initial particle distribution is of the primary concern.
For that purpose, we note that the first order time-derivative of the pair-production rate
\begin{eqnarray}
X_{(pl)} (t) : &=& \langle 0_{(pl)}, t \vert \hat{a}^{\dagger}_{(pl)} (- \infty) \hat{a}_{(pl)} (- \infty)
+ \hat{b}_{(pl)} (- \infty) \hat{b}^{\dagger}_{(pl)} (- \infty) \vert 0_{(pl)}, t \rangle \nonumber\\
&=& 1 + 2 {\cal N}_{(pl)} (t) \label{X}
\end{eqnarray}
is related with the other correlation functions
\begin{eqnarray}
Y_{(pl)} (t) : &=& i \langle 0_{(pl)}, t \vert \hat{a}_{(pl)} (- \infty) \hat{b}_{(pl)} (- \infty)
- \hat{a}^{\dagger}_{(pl)} (- \infty) \hat{b}^{\dagger}_{(pl)} (- \infty) \vert 0_{(pl)}, t \rangle,\nonumber\\
Z_{(pl)} (t) : &=& \langle 0_{(pl)}, t \vert \hat{a}_{(pl)} (- \infty) \hat{b}_{(pl)} (- \infty)
+ \hat{a}^{\dagger}_{(pl)} (- \infty) \hat{b}^{\dagger}_{(pl)} (- \infty) \vert 0_{(pl)}, t \rangle.
\end{eqnarray}
The unity in eq. (\ref{X}) is the zero-point energy of quantized field, which relates to
renormalization of the vacuum energy.
In fact, the quantum master equation consists of coupled linear differential equations \cite{KimSchubert}
\begin{eqnarray}
\dot{X}_{(pl)} &=& \frac{p(p+1)}{l \cosh^2 t} Y_{(pl)},\nonumber\\
\dot{Y}_{(pl)} &=& \frac{p(p+1)}{l \cosh^2 t} X_{(pl)} + \Bigl(2l + \frac{p(p+1)}{l \cosh^2 t} \Bigr) Z_{(pl)},\nonumber\\
\dot{Z}_{(pl)} &=& - \Bigl(2l + \frac{p(p+1)}{l \cosh^2 t} \Bigr) Y_{(pl)}. \label{master-eq}
\end{eqnarray}
The initial conditions are such that no pair is present and the correlation functions vanish
in the asymptotic region $t = - \infty$
\begin{eqnarray}
X (- \infty) = 1, \quad Y (- \infty) = Z (- \infty) = 0, \nonumber\\
\dot{X} (- \infty) = \dot{Y} (- \infty) = \dot{Z} (- \infty) = 0.
\end{eqnarray}

The quantum master equation (\ref{master-eq}) is equivalent to the linear equation \cite{KimSchubert}
\begin{eqnarray}
\frac{d^3 F_{(pl)}}{dt^3}  + 4\Bigl(l^2 + \frac{p(p+1)}{\cosh^2 t} \Bigr) \frac{d F_{(pl)}}{dt} + 2
\Bigl( \frac{d}{dt} \frac{p(p+1)}{\cosh^2 t} \Bigr) F_{(pl)}
 -  \frac{d}{dt} \frac{p(p+1)}{l^2\cosh^2 t} = 0, \label{lin-eq}
\end{eqnarray}
where
\begin{eqnarray}
F_{(pl)} (t) = \frac{1}{l} \int_{- \infty}^{t} d \xi Y_{(pl)} (\xi).
\end{eqnarray}
The pair-production rate is then given by
\begin{eqnarray}
X_{(pl)} (t) = 1 + \int_{- \infty}^t d \xi \dot{F}_{(pl)} (\xi) \frac{p(p+1)}{\cosh^2 \xi},
\label{pair-f}
\end{eqnarray}
and the other correlation functions are given by
\begin{eqnarray}
Y_{(pl)} (t) &=& l \dot{F}_{(pl)} (t), \nonumber\\
Z_{(pl)} (t) &=& - \int_{- \infty}^t d \xi \dot{F}_{(pl)} (\xi) \Bigl(2 l^2 + \frac{p(p+1)}{\cosh^2 \xi} \Bigr).
\label{int-cor}
\end{eqnarray}
Closely inspecting eqs. (\ref{pair-f}) and (\ref{int-cor}) and using another relation \cite{KimSchubert}
\begin{eqnarray}
\frac{d^2 F}{dt^2} (t) = \frac{\omega^2_{(pl)} (t) - l^2}{l^2} - 2 \omega^2_{(pl)}(t) F (t) - 2 \int_{- \infty}^t d \xi\dot{F} (\xi) \omega^2_{(pl)} (\xi)
\end{eqnarray}
leads to the conserved quantity for the solution $F$ to eq. (\ref{lin-eq})
\begin{eqnarray}
X^2 (t) - Y^2 (t) - Z^2 (t) = 1.
\end{eqnarray}

The particular form of eqs. (\ref{master-eq}) or (\ref{lin-eq}) or (\ref{pair-f}) and (\ref{int-cor}) suggests
the power series solution
\begin{eqnarray}
F_{(pl)} (t) = \sum_{k = 1} C_{(pl)k} \Bigl(\frac{1}{\cosh^2 t} \Bigr)^k, \label{F-ser}
\end{eqnarray}
and
\begin{eqnarray}
X_{(nl)} (t) = \sum_{k = 0} D_{(nl)k} \Bigl(\frac{1}{\cosh^2 t} \Bigr)^k, \quad D_{(nl)0} = 1. \label{X-ser}
\end{eqnarray}
The recurrence relation for $C_{(pl)k}$ can be found, for instance, from eq. (\ref{lin-eq})
\begin{eqnarray}
C_{(pl) k+1} &=& - \frac{(2k+1)(p-k)(p+k+1)}{2(k+1) ((k+1)^2 +l^2)} C_{(pl)k},\nonumber\\
C_{(pl)1} &=& \frac{p(p+1)}{4l^2 (1+ l^2)}, \label{F-coef}
\end{eqnarray}
and the coefficients $ D_{(pl)k}$ are found from eq. (\ref{pair-f})
\begin{eqnarray}
D_{(pl)k} = \frac{p(p+1)k}{k+1} C_{(pl)k}.
\end{eqnarray}
The coefficient $C_{(pl)k}$ may be written in a compact form
\begin{eqnarray}
C_{(pl) k} = (-1)^{k+1} \frac{(2k)!}{2^{2k+1} (k!)^2} \frac{1}{p(p-k)} \prod_{m = 0}^{k} \Bigl(\frac{p^2 - m^2}{l^2 + m^2} \Bigr).
\label{Fcoef-gen}
\end{eqnarray}

\section{Soliton Solutions by Algebraic Method}\label{gen-sol}

In this section we directly find the solition solutions by an algebraic method. We shall confine our study to the
gauge field leading to the squared energy
\begin{eqnarray}
\omega^2 (t) = \omega_0^2 (l^2 + U_0 u(t)),
\end{eqnarray}
where the profile function $u(t)$ has a finite integral over $(- \infty, \infty)$, and $u(t)$
and its derivative $\dot{u}(t)$ approach zero sufficiently rapidly
in the asymptotic limits $t = \pm \infty$. And we further assume that $u(t)$ has the following
property
\begin{eqnarray}
\dot{u}^2 (t) = \sum_{k = 1} {\cal D}_{k} u^k (t). \label{quad}
\end{eqnarray}
There is no constant term since $u (\pm \infty) = \dot{u} (\pm \infty) = 0$. The solitonic gauge field $u = 1/\cosh^2 (\omega_0t)$
has the property (\ref{quad}). In the rescaled time
$t := \omega_0 t$, the quantum master linear equation (\ref{lin-eq}) reads
\begin{eqnarray}
\frac{d^3 {\cal F}}{dt^3}  + 4(l^2 + U_0 u) \frac{d {\cal F}}{dt} + 2 U_0
\frac{d u}{dt} {\cal F} -  \frac{U_0}{l^2} \frac{du}{dt} = 0. \label{gen-lin-eq}
\end{eqnarray}

We search for the polynomial solution of the form
\begin{eqnarray}
{\cal F} (t) = \sum_{k =1}^{N}  {\cal C}_k u^k (t), \label{poly-sol}
\end{eqnarray}
which shares the same solitonic characteristics as $u$.
In order for the series solution
to be a polynomial for some finite $N$, the negative contribution to ${\cal C}_k$ from $d^3{\cal F}/dt^3$ should cancel the positive contributions
from $u d{\cal F}/dt$ and $u{\cal F}$, otherwise the recurrence relation will have
all positive coefficients and will become an infinite series.
For this reason, we assume an ansatz for the gauge field
\begin{eqnarray}
\dot{u}^2 = \alpha u^2 + \beta u^3, \quad (\alpha > 0), \label{2-rel}
\end{eqnarray}
which has the soliton solution
\begin{eqnarray}
u (t) = - \frac{\alpha}{\beta \cosh^2 (\frac{\sqrt{\alpha}t}{2})}.
\end{eqnarray}Furthermore,
the master linear equation (\ref{gen-lin-eq}) together with the ansatz (\ref{2-rel})
leads to the simplest linear recurrence relation
\begin{eqnarray}
{\cal C}_{k+1} &=& - \frac{(2k+1)[\beta k (k+1) + 4U_0]}{2 (k+1) [ \alpha (k+1)^2 + 4l^2]} {\cal C}_k, \nonumber\\
{\cal C}_1 &=& \frac{U_0}{l^2 (\alpha + 4 l^2)}. \label{gen-coef}
\end{eqnarray}
The series terminates when $U_0/\beta = - N(N+1)/4$ for some natural number $N$.
We recover the solitonic gauge fields in sec. \ref{sol-non-sol} by choosing $\alpha = 4$ and $\beta = -4$,
and the recurrence relation (\ref{gen-coef}) reduces to (\ref{F-coef}) for the natural number $p = N$ for
soliton solutions in sec. \ref{ser-sol}.

On the other hand, another simple ansatz, for instance, $\dot{u}^2 = \alpha u + \beta u^2$,
leads to a sinusoidal gauge potential, not a solitonic gauge field. Still another ansatz
\begin{eqnarray}
\dot{u}^2 = \alpha u^3 + \beta u^4
\end{eqnarray}
leads to a rational soliton solution \cite{DrazinJohnson}
\begin{eqnarray}
u = \frac{4 \alpha}{(\alpha t)^2 - 4 \beta}, \quad (\beta < 0), \label{rat-sol}
\end{eqnarray}
but the recurrence relation involves three terms
\begin{eqnarray}
{\cal C}_{k+1} &=& - \frac{(2k+1)[\alpha k (k+1) + 4U_0]}{8 l^2 (k+1)} {\cal C}_k
- \frac{\beta}{4 l^2}  (k-1) k {\cal C}_{k-1}, \nonumber\\
{\cal C}_1 &=& \frac{U_0}{4 l^4}. \label{rat-sol-coef}
\end{eqnarray}
The finite series requires $\beta = 0$ and $U_0/\alpha = - N(N+1)/4$ for some natural number $N$,
which makes the solution (\ref{rat-sol}) ill behaved near $t = 0$. Solving (\ref{rat-sol-coef})
and summing the series (\ref{poly-sol}) to get a closed form is beyond the scope of this paper,
which will be left as an open problem.

\section{Pair Production in Solitonic Gauge Fields}\label{pair-sol}

\begin{figure}[t]
%\subfloat[]
{\includegraphics[width=0.45\linewidth,height=0.25\textheight ]
{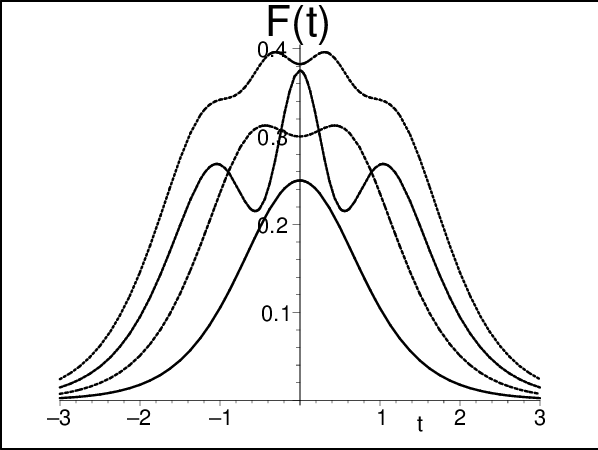} }\hfill
%\subfloat[]
{\includegraphics[width=0.45\linewidth,height=0.25\textheight ]
{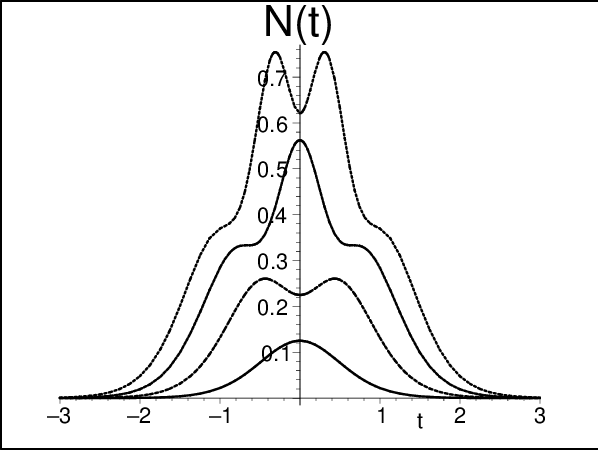} } \caption{The functions $F_{(n1)}$ and the pair-production rates ${\cal N}_{(n1)}$ are plotted against
$t(:= \omega_0 t)$: the solid line for $F_{(11)}$, the dotted line for $F_{(21)}$, the dashed line for $F_{(31)}$, and
the dash-dotted line for $F_{(41)}$ [left panel];
the solid line for ${\cal N}_{(11)}$, the dotted line for ${\cal N}_{(21)}$, the dashed line for ${\cal N}_{(31)}$, and
the dash-dotted line for ${\cal N}_{(41)}$ [right panel].} \label{fig1}
\end{figure}

In this section we consider (i) the solitonic gauge fields (\ref{sol-omega}), where $p = n$ is a natural number.
The solitonic gauge field corresponds to the reflectionless
scattering of the incoming positive frequency solution and predicts no pair production in the asymptotic limit $t = \infty$.
The series (\ref{F-ser}) becomes a polynomial solution
\begin{eqnarray}
F_{(nl)} (t) = \sum_{k = 1}^{n} C_{(nl)k} \Bigl(\frac{1}{\cosh^2 t} \Bigr)^k, \quad C_{(nl)1} &=& \frac{n(n+1)}{4l^2 (1+ l^2)} \label{sol-ser}
\end{eqnarray}
since $C_{(nl) k} = 0$ for $k \geq n+1$. Note that the time is rescaled as $t:= \omega_0t$. The coefficients change signs alternatively and the solution exponentially
decreases in the asymptotic regions $(t = \pm \infty)$, exhibiting solitonic characteristics. In fact, the solitonic solution
is a finite expression for the entire period for another exact formula (\ref{pair-mod-sol}) from the mode solution, which is
an infinite series. Though eq. (\ref{pair-mod-sol}) is equal to eq. (\ref{sol-ser}), any transformation formula
of the hypergeometric function is not available to our knowledge \cite{gr-table,as-table}. The solitonic solutions provide the global kinematics
for the pair-production rate and show the temporal nature of pair production.
The first four solutions of $F_{(nl)}$ for $l =1$ are given by
\begin{eqnarray}
F_{(11)}(t) &=& \frac{1}{4 \cosh^2 t}, \label{F-1} \\
F_{(21)}(t) &=& \frac{3}{4 \cosh^2 t} - \frac{9}{20 \cosh^4 t}, \label{F-2} \\
F_{(31)}(t) &=& \frac{3}{2 \cosh^2 t} - \frac{9}{4 \cosh^4 t} + \frac{9}{8 \cosh^6 t}, \label{F-3} \\
F_{(41)}(t) &=& \frac{5}{2 \cosh^2 t} - \frac{27}{4 \cosh^4 t} + \frac{63}{8 \cosh^6 t} - \frac{441}{136 \cosh^8 t}. \label{F-4}
\end{eqnarray}
The one-soliton solution (\ref{F-1}) is the same as eq. (25) of ref. \cite{KimSchubert}. The profiles are relatively simple
and their variations depend on quantum number $n$, as shown in fig. \ref{fig1}.

\begin{figure}[t]
%\subfloat[]
{\includegraphics[width=0.45\linewidth,height=0.25\textheight ]
{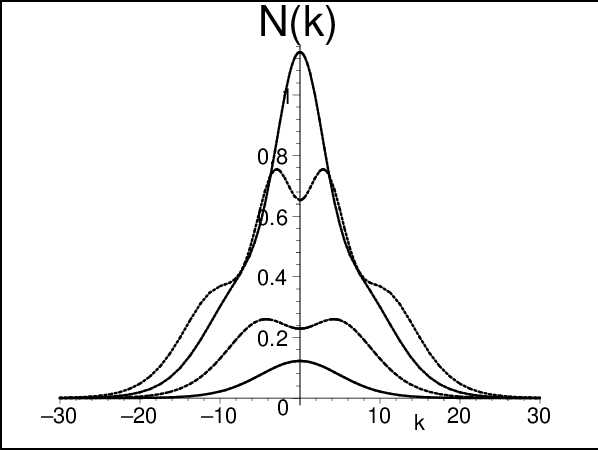} }\hfill
%\subfloat[]
{\includegraphics[width=0.45\linewidth,height=0.25\textheight ]
{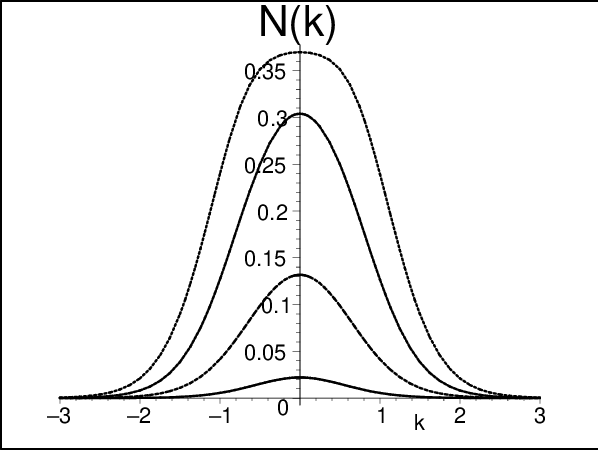} } \caption{The momentum distributions ${\cal N}_{(n1)} (k)$ are plotted against
$k(:=k/m)$: at $t(:=mt) =0.1$, the solid line for ${\cal N}_{(11)} (k)$, the dotted line for ${\cal N}_{(21)} (k)$, the dashed line for ${\cal N}_{(31)} (k)$, and
the dash-dotted line for ${\cal N}_{(41)} (k)$ [left panel]; at $t(:=mt) =1$, the solid line for ${\cal N}_{(11)}$, the dotted line for ${\cal N}_{(21)}$, the dashed line for ${\cal N}_{(31)}$, and
the dash-dotted line for ${\cal N}_{(41)}$ [right panel].} \label{fig2}
\end{figure}

\begin{figure}[t]
%\subfloat[]
{\includegraphics[width=0.45\linewidth,height=0.25\textheight ]
{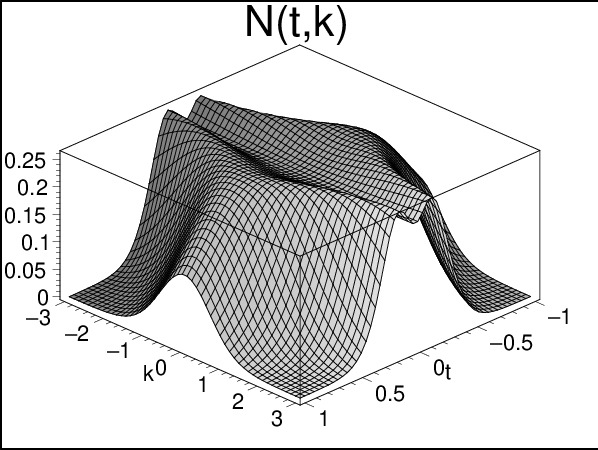} }\hfill
%\subfloat[]
{\includegraphics[width=0.45\linewidth,height=0.25\textheight ]
{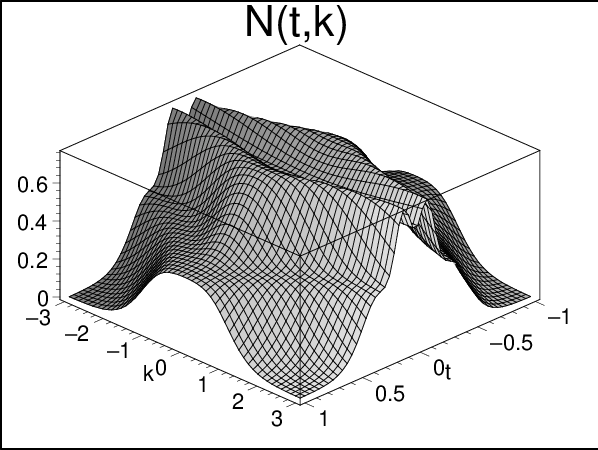} } \caption{The time-momentum distributions ${\cal N}_{(21)} (t,k)$ are plotted against
$t(:= m t)$ and $k(:=k/m)$ [left panel] and ${\cal N}_{(41)} (t,k)$ [right panel].} \label{fig3}
\end{figure}

From eq. (\ref{pair-f}) the pair-production rate is given by
\begin{eqnarray}
{\cal N}_{(nl)} (t) = \frac{n(n+1)}{2} \sum_{k = 1}^{n} \frac{k}{k+1} C_{(nl) k} \Bigl(\frac{1}{\cosh^2 t} \Bigr)^{k+1}.
\end{eqnarray}
The first term is the leading term (\ref{leading-pair}) from the mode solution in the asymptotic regions $(t = \pm \infty)$.
The pair-production rate in the first four solitonic gauge fields for $l=1$  are
\begin{eqnarray}
{\cal N}_{(11)}(t) &=& \frac{1}{8 \cosh^4 t}, \label{N-1} \\
{\cal N}_{(21)}(t) &=& \frac{9}{8 \cosh^4 t} - \frac{9}{10 \cosh^6 t}, \label{N-2} \\
{\cal N}_{(31)}(t) &=& \frac{9}{2 \cosh^4 t} - \frac{9}{\cosh^6 t} + \frac{81}{16 \cosh^8 t}, \label{N-3}\\
{\cal N}_{(41)}(t) &=& \frac{25}{2 \cosh^4 t} - \frac{45}{\cosh^6 t} + \frac{945}{16 \cosh^8 t} - \frac{441}{17\cosh^{10} t}. \label{N-4}
\end{eqnarray}
The pair-production rates in fig. \ref{fig1} exhibit the solitonic characteristics. They increase as
quantum number $n$, the strength of interaction.
The pair-production rate from the master equation exponentially decreases to zero, does not have
any oscillatory tailing behavior, and exhibits the solitonic characteristics. The momentum distribution
of the pair-production rate has a substructure for time less than one Compton time $t(:=mt) =1$, as shown in fig. \ref{fig2}.
And the substructure appears within the domain of a few Compton times and wavelengths larger than a few Compton wavelength,
as shown in fig. \ref{fig3}. In fact. the substructure depends on the combination of $\omega_0 t = \sqrt{1+ (k/m)^2} (mt)$.

\begin{figure}[t]
%\subfloat[]
{\includegraphics[width=0.45\linewidth,height=0.25\textheight ]
{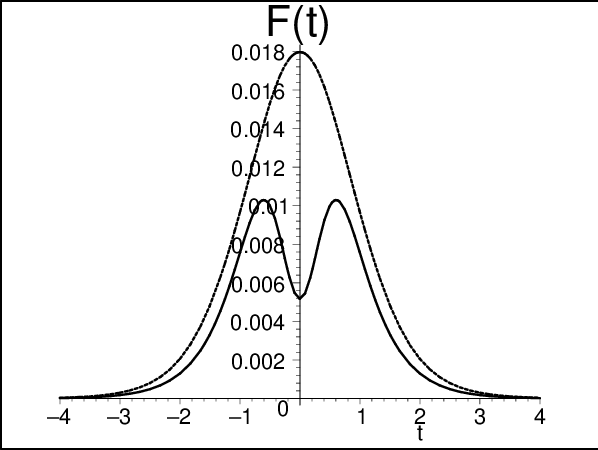} }\hfill
%\subfloat[]
{\includegraphics[width=0.45\linewidth,height=0.25\textheight ]
{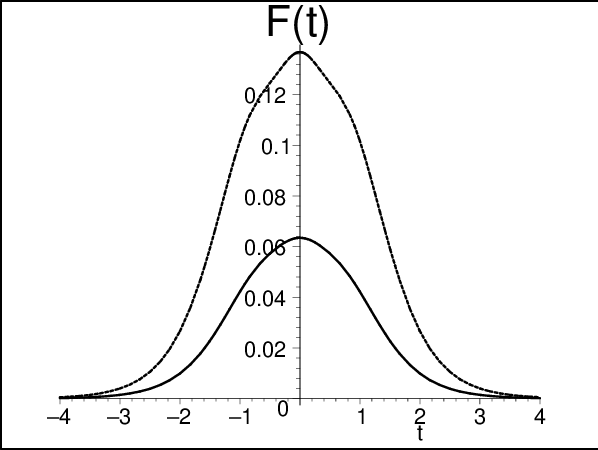} } \caption{The functions $F_{(nl)}$ are plotted against
$t(:= \omega_0 t)$: the solid line for $F_{(44)}$ and the dotted line for $F_{(4\frac{7}{2})}$ [left panel];
and the solid line for $F_{(32)}$ and the dotted line for $F_{(3\frac{3}{2})}$ [right panel].} \label{fig4}
\end{figure}

We now turn to the second case of (\ref{non-sol-omega}) with natural numbers $n$ and $l \geq 2$. The mode solution (\ref{mod-sol}) does
not have the negative frequency branch even for any $l$  at $t = \infty$ as shown in sec. \ref{sol-non-sol}.
The master linear equation (\ref{lin-eq}) still has the polynomial solutions (\ref{sol-ser}) with (\ref{Fcoef-gen}).
In fig. \ref{fig4} we compare $F_{(44)}$ with $F_{(4\frac{7}{2})}$ and also compare $F_{(32)}$ with $F_{(3\frac{3}{2})}$.
The non-natural number $l$ drastically changes the profiles of $F_{(4)}$ and $F_{(3)}$ with $l =1$ and other natural number $l$.

\section{Comparison with Non-Solitonic Gauge Fields}\label{non-sol-pair}

Finally we consider the mode equation in the non-solitonic gauge fields with positive non-natural number $p$
\begin{eqnarray}
\ddot{\varphi}_{(pl)} (t) + \Bigl(l^2 + \frac{p(p+1)}{\cosh^2 (t)} \Bigr)
\varphi_{(p)} (t) = 0. \label{nonsol-mod-eq}
\end{eqnarray}
One is tempted to use the series solution (\ref{F-ser}), which becomes an infinite power series since
the coefficients do not terminate for non-natural number $p$.
The series solution (\ref{F-ser}) can be written as
\begin{eqnarray}
F_{(pl)} (t) = \sum_{k = 1}^{[p]} C_{(pl)k} \Bigl(\frac{1}{\cosh^2 t} \Bigr)^k + \sum_{k = [p]+1}^{\infty} C_{(pl)k} \Bigl(\frac{1}{\cosh^2 t} \Bigr)^k, \label{non-sol-F}
\end{eqnarray}
where the first sum is finite and alternating while the second sum is non-alternating and infinite.
The series (\ref{non-sol-F}) converges strongly near the limit $t = - \infty$ and gives an accurate result.
The first term is the leading term for the pair-production rate (\ref{leading-pair}) from the mode solution.
As the coefficients (\ref{Fcoef-gen})
do not change the sign and approximately approach a constant value $C_{N}$ times $1/k$ for $k \geq N$ for some large $N$,
the series solution for large $k$ converges asymptotically to a logarithm
\begin{eqnarray}
F_{(pl)} (t) \approx - C_{N} \ln \Bigl(1- \frac{1}{\cosh^2 t} \Bigr) \Bigl(\frac{1}{\cosh^2 t} \Bigr)^N.
\end{eqnarray}
Therefore the series solution cannot be relied on when $\cosh t \approx 1$ and near $t = 0$. This implies
that the formula (\ref{pair-f}) for pair-production rate may not be used beyond the limit in which the series (\ref{non-sol-F})
converges since it is not justified to evolve (\ref{non-sol-F}) from one asymptotic limit $t = - \infty$ to the other limit
$t = \infty$ through $t=0$. Thus the series solution to the master equation does not provide useful information
on the global kinematics for pair production beyond $t = 0$ more than the mode solution, as shown below.

The exact solution (\ref{mod-sol}), which is valid for the entire period, can be transformed into another form useful
in the other asymptotic limit $t = \infty$
\begin{eqnarray}
\varphi_{(pl)} (t) &=& \frac{e^{-il t}}{\sqrt{2}} e^{i \vartheta_{\Gamma}} \Bigl(\frac{\sin^2 (\pi p) \cosh^2 (\pi l)+
\cos^2 (\pi p) \sinh^2 (\pi l) }{\sinh^2 (\pi l)} \Bigr)^{1/2} B_{(pl)} (t)
\nonumber\\
&& + \frac{e^{il t}}{\sqrt{2}} e^{i \frac{\pi}{2}} \Bigl(\frac{\sin (\pi p)}{\sinh (\pi l)} \Bigr) B_{(pl)}^* (t),
\end{eqnarray}
where $\vartheta_{\Gamma}$ is the phase of the gamma functions of (\ref{asym-sol2}) and $B_{(pl)}(t)$ is another
time-dependent amplitude
\begin{eqnarray}
B_{(pl)}(t) = (1+ e^{-2t})^{p+1} F(p+1, p+1+il; 1+il; -e^{-2t}).
\end{eqnarray}
A straightforward calculation gives the first two leading terms of the pair-production rate
\begin{eqnarray}
{\cal N}_{(pl)} (t) &=& \Bigl(\frac{\sin (\pi p)}{\sinh (\pi l)} \Bigr)^2 \Bigl(1 - 2 \frac{p(p+1)}{1+l^2} e^{-2 t} \Bigr)
- 2 \frac{p(p+1)}{(1+l^2)^{1/2}} \Bigl(\frac{\sin (\pi p)}{\sinh (\pi l)} \Bigr)
\nonumber\\&& \times \Bigl(\frac{\sin^2 (\pi p) \cosh^2 (\pi l)+
\cos^2 (\pi p) \sinh^2 (\pi l) }{\sinh^2 (\pi l)} \Bigr)^{1/2} \sin(lt) \sin(lt + \frac{\pi}{2} - \vartheta_{\Gamma} + \vartheta_l)
e^{-2t},\nonumber\\
\label{non-sol-osc}
\end{eqnarray}
where
\begin{eqnarray}
\sin \vartheta_{l} = \frac{l}{(1+l^2)^{1/2}}, \quad \cos \vartheta_{l} = \frac{1}{(1+l^2)^{1/2}}.
\end{eqnarray}
The frequency of $2l$ in the oscillatory factor of the second term is predicted by the non-adiabatic
and the adiabatic quantum master equations (\ref{mas-eq}) and (\ref{ad-mas-eq}). For a natural number $p = n$, the rate (\ref{non-sol-osc})
vanishes as expected.

\begin{figure}[t]
%\subfloat[]
{\includegraphics[width=0.45\linewidth,height=0.25\textheight ]
{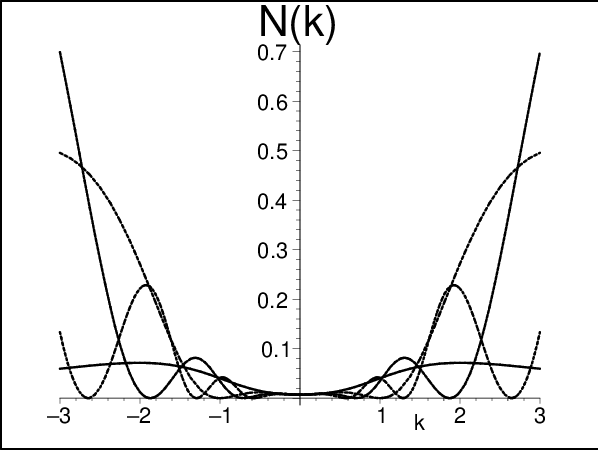} }\hfill
%\subfloat[]
{\includegraphics[width=0.45\linewidth,height=0.25\textheight ]
{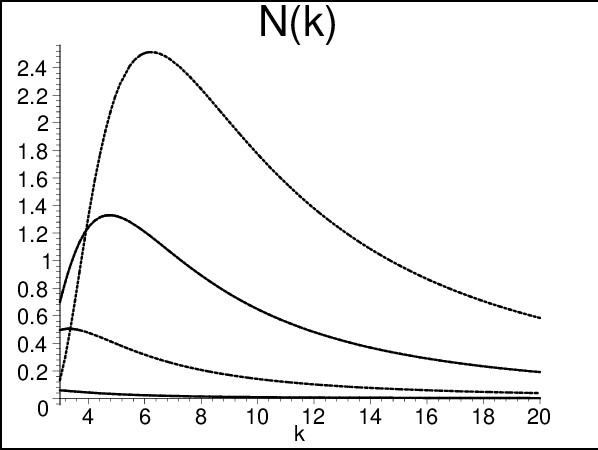} } \caption{The momentum distributions of ${\cal N}(k) = (\sin \pi p/\sinh \pi l)^2$ for $V_0/m = 1$ against
$k(:= k/m)$: the solid line for $U_0/m^2 =1$, the dotted line for $U_0/m^2 =4$, the dashed line for $U_0/m^2 =9$, and
the dash-dotted line for $U_0/m^2 =16$ in the range of $-3 \leq k \leq 3$ [left panel]
and in the range of $3 \leq k \leq 30$ [right panel].} \label{fig5}
\end{figure}

For the squared energy of the form
\begin{eqnarray}
\omega^2 (t) = V_0^2 + \frac{U_0}{\cosh^2 (\omega_0 t)},
\end{eqnarray}
where $U_0$ is an interaction strength, we identify the parameters
\begin{eqnarray}
l = \frac{V_0}{\sqrt{m^2 + {\bf k}^2}}, \quad p = \sqrt{\frac{U_0}{m^2 + {\bf k}^2} + \frac{1}{4}} - \frac{1}{2}.
\end{eqnarray}
Noting again $t:=\omega_0 t$ and $\omega_0 = \sqrt{m^2 + {\bf k}^2}$, the constant term in eq. (\ref{non-sol-osc}) depends on
the momentum through the sinusoidal function in a nontrivial way. Fig. \ref{fig5} shows the momentum distribution
of the pair-production rate, which has a substructure for small momentum but exponentially suppressed for
high momentum. Furthermore, the decaying terms also depend on
the momentum both through the amplitudes and through the time-dependence.
The rich substructure of momentum distribution distinguishes the non-solitonic gauge fields from the solitonic gauge fields.

\section{Back-Reaction Problem}\label{back-reaction}

In this section we investigate the back-reaction of produced pairs on the solitonic fields.
The current of a charged scalar particle in an electromagnetic field is the expectation value
with respect to the vacuum with quantum numbers $n$ and $l$
\begin{eqnarray}
\langle \hat{J}_{\parallel} (t) \rangle &=& 2 q \int \frac{d^3 {\bf k}}{(2 \pi)^3} \frac{k_{\parallel} - q A_{\parallel}}{\omega_{\bf k} (0)}
\Bigl(\frac{1}{2} + {\cal N} (t) + Z(t) \Bigr) \nonumber\\
&=& \frac{2 q}{l} \int \frac{d^3 {\bf k}}{(2 \pi)^3} \sqrt{k_{\parallel}^2 + \frac{n(n+1)}{\cosh^2 (\omega_0 t)}}
\Bigl(\frac{1}{2} + {\cal N} (t) + Z(t) \Bigr). \label{back-r}
\end{eqnarray}
The first term is the vacuum energy of the quantum field, which has to be regulated away through renormalization,
and the pair-production rate ${\cal N} (t)$ and the correlation function $Z(t)$ of eq. (\ref{int-cor}) are finite
for solitonic gauge fields as shown in sec. \ref{pair-sol}.
To understand the time-dependent of the induced current,
we consider the simplest case ${\bf k} = 0$ for which $\omega_0 = m$ and $t:= mt$.
The induced current (\ref{back-r}) leads to a linear equation for the induced electric field
\begin{eqnarray}
\dot{E}_{\rm in} (t) = - \langle \hat{J}_{\parallel} (t) \rangle = - \frac{2 q\sqrt{n(n+1)} }{l}\frac{1}{\cosh t} \Bigl(\frac{1}{2} + {\cal N} (t) + Z(t) \Bigr).
\end{eqnarray}
The induced current, for instance, for one-soliton $(n=1)$ gauge field for arbitrary $l$ is
\begin{eqnarray}
E_{\rm in} (t) &=& \frac{2\sqrt{2} q }{l} \Biggl[ \frac{1}{2(1+l^2)} \Bigl(1 + \frac{3}{16 l^2} \Bigr) \frac{\sinh t}{\cosh^2 t}
+ \frac{1}{16 l^2(1+l^2)} \frac{\sinh t}{\cosh^4 t} \nonumber\\ && -
\frac{1}{2} \Bigl(1 - \frac{1}{1+l^2} - \frac{1}{16 l^2(1+l^2)} \Bigr) \arctan (\sinh t) \Biggr]. \label{ind-E}
\end{eqnarray}
As $\arctan (\sinh t) = \pm \pi/2$ at $t = \pm \infty$, we regulate away the last term through
renormalization of the vacuum energy and the charge.
Then the ratio of the renormalized induced current to the solitonic field background is
\begin{eqnarray}
\frac{E^{\rm ren}_{\rm in} (t)}{E_{\rm ex} (t)} = - \frac{q^2}{l(1+l^2)} \Bigl( 1 + \frac{3}{16 l^2} + \frac{1}{8l^2}
\frac{1}{\cosh^2 t} \Bigr),
\end{eqnarray}
and for the electron-positron pairs is order of the fine structure constant $(e^2)$ in the units of $c = \hbar =1$.
Therefore the back-reaction is small compared with the background solitonic gauge field.
Including the momentum integration, the rate of the renormalized induced electric field for one-soliton gauge field is
\begin{eqnarray}
\dot{E}^{\rm ren}_{\rm in} (t) = \frac{2 q}{l(1+l^2)} \int \frac{d^3 {\bf k}}{(2 \pi)^3} \sqrt{k_{\parallel}^2 + \frac{n(n+1)}{\cosh^2 (\omega_0 t)}}
\Bigl(\frac{1}{\cosh^2 (\omega_0 t)} + \frac{1}{4l^2 \cosh^2 (\omega_0 t)} - \cdots \Bigr), \label{back-renor}
\end{eqnarray}
where the dots denote a subtraction scheme for renormalization. We argue that the back-reaction of produced pairs
is smaller by the fine structure constant than the solitonic gauge field background and can be safely neglected.
It remains open to show the existence of solitonic gauge fields including the back-reaction, which is the beyond the scope
of this paper.

\section{Discussion}\label{dis}

One may ask the physical meaning of the pair-production rate during the
 interaction with the electric field. It is defined either by the number of pairs carried by the
 time-dependent vacuum as measured by the in-vacuum number operator or by the number of pairs
 contained in the in-vacuum as measured by the time-dependent number operator, both of which
 give the identical result because of the Bogoliubov transformation and the inverse transformation (\ref{bog-tr}).
 In the in-out formalism, the pair-production rate is determined by the Wronskian (\ref{bog-co}) between
 the mode solution for the in-vacuum and the mode solution at arbitrary time. Does the pair-production rate
 refer to real pairs to be measured by a detector?
In the solitonic gauge fields
 in which the pair-production rate exponentially increases from and then decreases to zero in a symmetric way,
 do the pairs produced during the evolution really annihilate themselves in the end? If not, where do the pairs
 disappear?
This question is rooted on the fundamental question for the particle concept when
its quantum field depends on time in a nontrivial way, which has been frequently raised for
expanding spacetimes.

\begin{figure}[t]
%\subfloat[]
{\includegraphics[width=0.45\linewidth,height=0.25\textheight ]
{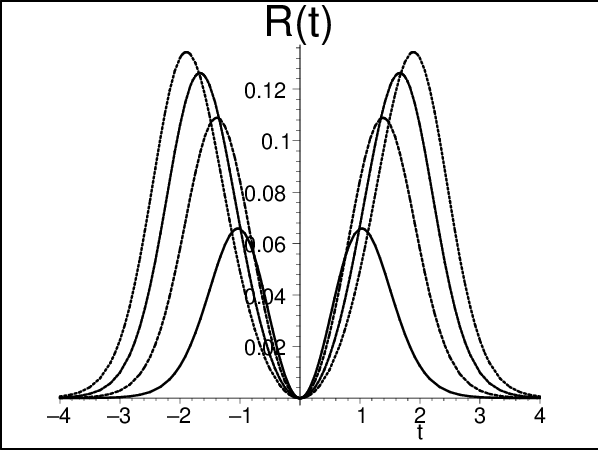} }\hfill
%\subfloat[]
{\includegraphics[width=0.45\linewidth,height=0.25\textheight ]
{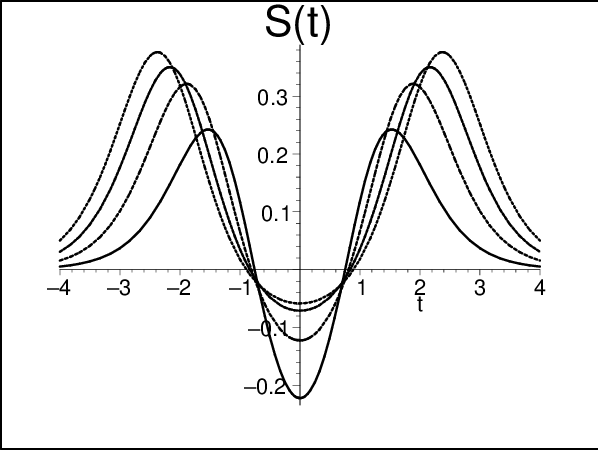} } \caption{The adiabaticity parameter $R(t) := (\dot{\omega})^2_{(nl)}/\omega^4_{(n1)}$
is plotted against $t (:=\omega_0 t)$: the solid line for $n=1$, the dotted line for $n=2$, the dashed line for $n=3$, and
the dash-dotted line for $n=4$ [left panel]. And another adiabaticity parameter $S(t) := \ddot{\omega}_{n1}/\omega^3_{(n1)}$ is plotted:
the solid line for $n=1$, the dotted line for $n=2$, the dashed line for $n=3$, and
the dash-dotted line for $n=4$ [right panel].} \label{fig6}
\end{figure}

It is not the purpose of this paper to answer the fundamental question of what is the particle concept when
its quantum field varies in time. Instead, we pursue a pragmatic point of view by searching for regions in which
the particle concept and particle number makes a sense, at least in the adiabatic sense. In the regions in
which the adiabaticity holds, we may use the adiabatic mode functions and the adiabatic particle
number \cite{BirrelDavies}. In the case of solitonic gauge fields
the adiabaticity requires that
\begin{eqnarray}
\frac{\dot{\omega}^2_{(nl)}}{\omega^4_{(nl)}} &=& \frac{(n(n+1)\sinh t)^2}{(l^2 \cosh^2 t + n(n+1))^{3}}, \nonumber\\
\frac{\ddot{\omega}_{(nl)}}{\omega^3_{(nl)}} &=& \frac{n(n+1) \cosh^2 t}{(l^2 \cosh^2 t + n(n+1))^{3}} \Bigl[
n(n+1) \Bigl(1 - \frac{2}{\cosh^2 t} \Bigr) + l^2 (2 \cosh^2 t - 3) \Bigr]
\end{eqnarray}
should be smaller than unity. This condition is always satisfied except for the region $|t| \leq 4$
for small natural number $n$. Even in this limited region
the ratios are still less than $1/2$ and the adiabaticity holds in a broad sense, as shown in fig. \ref{fig6}.
In particular, the ratios are exponentially
suppressed in the asymptotical regions $|t| \gg 1$ and particles are well-defined there.
The larger quantum number $n$ corresponds to higher energy than the smaller quantum number
and thus suppresses further the ratios. The pair-production rate begins from the
region $t = - \infty$ in which the particle number is well-defined,
passes a region around $t = 0$ in which the adiabaticity roughly holds, and finally settles down
in the region $t = \infty$. Therefore we may interpret the pair-production rate as the adiabatic particle
pairs or in the literal sense as the exact correlation function for the number operator
that extrapolates the two asymptotic regions.

It now remains the question of how the pairs produced during the interaction behave and disappear such that
the pair-production rate vanishes in the other asymptotic limit.
In the solitonic gauge fields pairs are effectively produced
roughly within the interval $|t| \leq 4$, i.e., eight Compton times (see fig. \ref{fig1}).
During the first
four Compton times particles accelerate to the positive direction and antiparticles to the negative direction
whereas during the next four Compton times the directions of particles and antiparticles
reverse because of the polarization change of the electric field. The pairs that are produced during $ 1 \leq |t| \leq 4$
may be regarded as real entities since they have enough time to separate over one Compton wavelength. However,
the pairs produced during $|t| \leq 1$ are likely to cancel each other due to the polarity change, in a similar
way as the virtual pair production and subsequent annihilation. Without any detection, the particles
keep quantum coherence with antiparticles and annihilate during the opposite polarity of electric field.
However, any detector to measure the pairs may break the quantum coherence and the pairs may not
completely cancel each other. To install a detector, one has to include an interaction between the detector and
the charged field, which makes an interaction theory and goes beyond the scope of this paper.

Finally, we compare the solitonic gauge fields with other temporally localized electric fields.
The frequently studied model is the Sauter electric field
\begin{eqnarray}
E_{\parallel} (t) = \frac{E_0}{\cosh^2 (t/\tau)}
\end{eqnarray}
with the gauge field
\begin{eqnarray}
A_{\parallel} (t) = - E_0 \tau ( \tanh (t/\tau) + 1). \label{sauter}
\end{eqnarray}
The electric field acts effectively during the period $|t| \leq 3 \tau$ and does not change the polarity.
Hence a charged particle gains the potential difference of $2 q E_0 \tau$ between two asymptotic regions
and the out-vacuum differs from the in-vacuum, implying pair production \cite{KimPage07,DunneHall}.
In fact, the Sauter gauge field (\ref{sauter})
does not satisfy the condition $A_{\parallel} (\infty) = 0$ for solitonic gauge fields
in sec. \ref{gen-sol} which is needed  for finiteness of pairs produced for the entire period
and other characteristics. Another localized electric field
\begin{eqnarray}
E_{\parallel} (t) = - \frac{E_0 \sinh t}{\cosh^2 t}
\end{eqnarray}
and its gauge field
\begin{eqnarray}
A_{\parallel} (t) = - \frac{E_0}{\cosh t} \label{local}
\end{eqnarray}
shares some property with the solitonic gauge field. The squared energy
\begin{eqnarray}
\omega^2 (t) = \Bigl(k_{\parallel} + \frac{qE_0}{\cosh t} \Bigr)^2 + {\bf k}^2_{\perp} + m^2
\end{eqnarray}
in terms of the profile function $u = 1/\cosh t$ can be transformed into the relation (\ref{quad}),
but the recurrence relation involve more than three terms, which does not necessarily
guarantee solitonic solutions, except for the zero longitudinal momentum.

\section{Conclusion}\label{con}

We have studied Schwinger pair production in gauge fields that lead to the energy of P\"{o}sch-Teller potential
(\ref{posch-teller}) for a charged boson. The gauge fields are classified into two categories: the solitonic gauge fields with
natural numbers and the non-solitonic gauge fields with non-natural numbers.
The pair-production rate from the mode solution in a solitonic gauge field exponentially decays
in the asymptotic regions while a non-solitonic gauge field leads to a constant pair-production rate
in addition to exponentially decaying terms in the future infinity.
Solving the quantum master equation, we find the exact pair-production rate in the solitonic gauge field which is a
polynomial of inverse square of hyperbolic cosine and exponentially decays in a symmetric way
in the asymptotic regions. Further, the pair-production rate in the solitonic gauge field
exhibits a non-oscillatory tailing behavior and a simple momentum distribution through the entire period.
None of these characteristics is shared by the non-solitonic gauge fields.
The exact information of pair production in solitonic gauge fields may shed light on understanding how the vacuum
structure changes and the produced pairs behave during the interaction with electromagnetic fields.

Schwinger pair production in time-dependent electric fields can be obtained either by
finding the Bogoliubov coefficients from the mode equation or by using the real-time propagator.
In this paper we have used the quantum master equation for the three correlation functions of quadratic
annihilation and creations operators: the number of pairs and the correlations for pair production
and pair annihilation. The quantum master equation allows us to directly integrate the pair-production
rate from gauge fields only and is easy to implement the condition for solitonic characteristics.
For that purpose, first we have extended the one-solitonic gauge field in ref. \cite{KimSchubert} to
multi-solitonic gauge fields of the same potential shape and then solved the quantum master equation
to find the pair-production rate with solitonic characteristics. Second, we have solved the quantum
master equation by an algebraic method and showed that the solitonic gauge fields (\ref{posch-teller})
with natural number $p$ are the simplest ones that satisfy the property imposed on the gauge fields. 
The algebraic method may open a possibility for other solitonic gauge fields.

In the gauge fields (\ref{posch-teller}) the pair-production rate has the meaning of adiabatic
particle-antiparticle pairs produced during the interaction of the electric field
since the mode equation satisfies the adiabaticity for the entire period as shown in sec. \ref{dis}.
However, it may be used even beyond the adiabatic limit as the exact correlation function that
extrapolates between asymptotic regions, in which particles are well-defined.
One interesting feature of the solitonic gauge fields is that the back-reaction from pairs
through the induced electric field is smaller than the background gauge fields by order of
the fine structure constant. Therefore, the back-reaction problem can be avoided
for the solitonic gauge fields. It remains an open question to show the existence of
solitonic gauge field including the back-reaction, for which the quantum master equation
for the correlation functions is now coupled to the induced electric field, which in turn coupled to
the correlation functions themselves.

\acknowledgments

The author thanks Christian Schubert for helpful discussions which have clarified some concepts of this paper
and for suggesting the momentum distribution,
and also thanks Hyun Kyu Lee for useful comments on the manuscript. He thanks Pauchy W-Y. Hwang, Don N. Page
and Yongsung Yoon for early collaborations on Schwinger pair production and QED effective actions. He appreciates
the warm hospitality at Instituto de F\'{i}sica y Matem\'{a}ticas, Universidad Michoacana de San Nicol\'{a}s de Hidalgo.
This work was supported in part by Basic Science Research Program through
the National Research Foundation of Korea (NRF) funded by the Ministry of Education, Science and Technology (2011-0002-520).

\end{document}